\documentclass[twocolumn]{emulateapj}
\usepackage{graphicx, amsmath, amsthm, amstext,mathtools,amssymb,color}

\usepackage{graphics}
\usepackage{natbib}
\usepackage{ulem}
\usepackage{relsize}

\usepackage[breaklinks,colorlinks,citecolor=blue,linkcolor=magenta]{hyperref} 
\usepackage[all]{hypcap} 

\def\out{_{\rm out}}

\def\i{_{\rm in}}
\def\out{_{\rm out}}

\def\ein{{\bf e}_{\rm in}}
\def\jin{{\bf j}_{\rm in}}
\def\jout{{\bf j}\out}

\newcommand{\ba}{\begin{eqnarray}}
\newcommand{\ea}{\end{eqnarray}}

\graphicspath{{./}{figures/}}

\shorttitle{A resonant origin for inclined planets}
\shortauthors{Petrovich et al.}

\begin{document}

\title{A disk-driven resonance as the origin of high inclinations of close-in planets}
\author{Cristobal Petrovich\altaffilmark{1}}
\author{Diego J. Mu\~noz\altaffilmark{2}}
\author{Kaitlin M. Kratter\altaffilmark{1}}
\author{Renu Malhotra\altaffilmark{3}}
\altaffiltext{1}{Steward Observatory, University of Arizona, 933 N. Cherry Ave., Tucson, AZ 85721, USA}
\altaffiltext{2}{CIERA, 
Northwestern University,
1800 Sherman Ave.,
Evanston, IL 60208, USA}
\altaffiltext{3}{Lunar and Planetary Laboratory, The University of Arizona, Tucson, AZ 85721, USA}

\begin{abstract}
The recent characterization of transiting close-in planets has revealed an intriguing population of sub-Neptunes with highly tilted and even polar orbits 
relative to their host star's equator.
Any viable theory for the origin of these close-in, polar planets must explain (1) the observed stellar obliquities, (2) the substantial eccentricities, and (3) the existence of Jovian companions with large mutual inclinations. In this work,
we propose a theoretical model that satisfies these requirements without invoking tidal dissipation or large primordial inclinations. Instead, tilting is facilitated by the protoplanetary disk dispersal during the late stage of planet formation, initiating a process of resonance sweeping and parametric instability. This mechanism consists of two steps. First, a nodal secular resonance excites the inclination to large values; then, once the inclination reaches a critical value, a linear eccentric instability is triggered, which 
detunes the resonance and ends inclination growth. The critical inclination is pushed to high values by general relativistic precession, making polar orbits an inherently post-Newtonian outcome. Our model predicts that polar, close-in sub-Neptunes coexist with cold Jupiters in low stellar obliquity orbits. 
\end{abstract}

\section{Introduction}

Although a large fraction of the multi-planet systems discovered by the
{\it Kepler} spacecraft exhibit a great
degree of coplanarity (\citealt{winn_review}),
some systems possess significant mutual inclinations \citep{MF2017,Zhu2018,pimen1},
pointing to unruly dynamical histories. 
Similarly, a large stellar obliquity --the tilt between the planet's orbital plane and the stellar equator-- can also indicate a period of dynamical upheaval.
Ensembles of obliquity measurements can
be used to probe the origin and dynamics of tilted systems
\citep[e.g.][]{fab09,mor14,mun18},  providing a powerful tool to study planet formation.

\begin{figure*}[t!]
\center
\includegraphics[width=18.5cm]{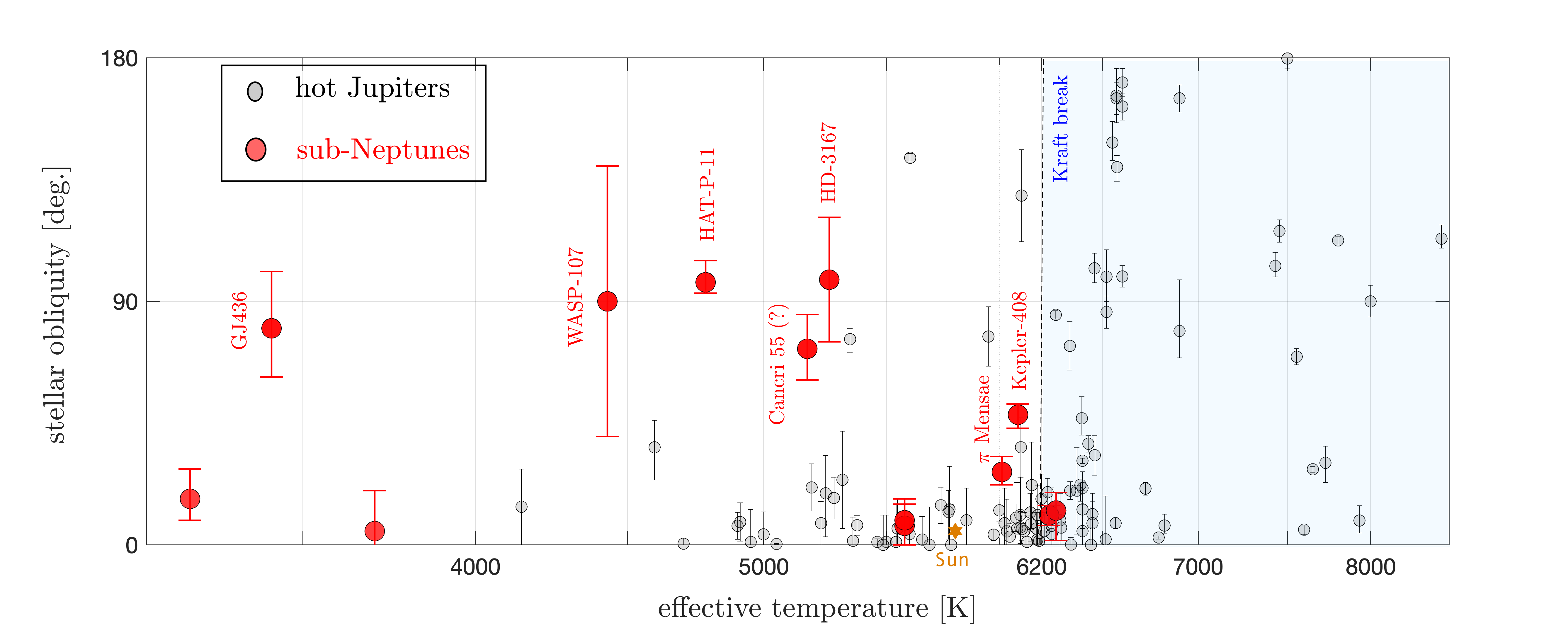}
\caption{Measured stellar obliquity for close-in planets ($a\lesssim0.1$ au) as a function of the  host star's effective temperature. The gray circles show the sample of hot Jupiters ($M_p>0.3 M_J$ and $P<10$ days) with reliable obliquity measurements (1-$\sigma$ errors $<20^\circ$). The larger red circles show the sample of planets with either sizes or masses comparable to or smaller than that of Neptune, specifically $R_p<6R_\oplus$ and/or $M_p<30 M_\oplus$.  The data is taken from the TEPCat Catalog as of August 2020 (\citealt{Southworth2011},     \href{http://www.astro.keele.ac.uk/jkt/tepcat/}{\color{blue}http://www.astro.keele.ac.uk/jkt/tepcat}) with most values corresponding to projected stellar obliquities, though a small fraction are non-projected values. When both are available, we use the latter.}	
\label{fig:psi_teff}
\end{figure*}

Owing to observational selection, most measurements of stellar obliquity have been made for hot Jupiter systems. 
Naturally, most theoretical efforts have focused
on explaining the obliquities of these systems.
Lower mass planets, however, are far more common than hot Jupiters (\citealt{winn_review}), and are less likely to realign the star via tidal interactions.
Consequently, smaller-mass planets offer a more representative and a more
pristine probe into the typical planet formation process.
Fortunately, modern instruments and novel analysis techniques are beginning
to provide obliquity measurements for planets in the sub-Neptune category.
 In Figure \ref{fig:psi_teff}, we display a subset of systems with obliquity measurements, highlighting 13 systems hosting sub-Neptunes, 5
of which are dramatically tilted into polar orbits.

Among the peculiarities of polar Neptunes, we highlight their propensity to have Jovian outer companions \citep{yee}, their non-negligible eccentricities \citep{correia}, and their occurrence in compact multi-planet systems \citep{dalal}. These properties limit the applicability of theoretical models developed to explain obliquities in hot Jupiters systems. For example, tilting the entire protoplanetary disk (e.g., \citealt{batygin12}) fails to explain why the inner planets in HAT-P-11 and $\pi$ Mensae have substantial mutual inclinations relative to their outer giant planet companions \citep{pimen1,gaia_inc2,gaia_inc3}, nor does it account for the significant eccentricities of close-in sub-Neptunes (e.g., HAT-P-11b has $e\simeq0.2$). The widely invoked mechanism of high-eccentricity migration that naturally leads to large obliquities of planets lacking nearby neighbors, is halted by the presence of other close-in planets \citep{mustill}, thus failing to explain polar compact multi-planet systems like HD-3167 \citep{dalal}. Moreover, the high-eccentricity migration hypothesis does not address the origin of the large initial inclinations required  for the mechanism to operate (e.g., $\gtrsim 70^\circ$ as proposed in GJ-436, \citealt{bourrier_gj}).

In this work, we propose a model that can explain eccentric, polar orbits of close-in planets that requires only the presence of an outer Jovian companion and a slowly decaying outer protoplanetary disk. As the disk decays, high stellar obliquities are generated via a two-step process: (1) a nonlinear secular resonance that excites orbital inclination and (2) saturation of inclination via a linear eccentric instability. This process produces highly inclined planets, often with eccentric orbits, and does not require extreme primordial inclinations of the planets or the disk.

\section{Two-Planet Systems with Dispersing Disks}
\label{sec:model}
Close-in planets ($a\i\lesssim 0.1$ au) are often accompanied by cold Jovians ($a\out\sim1-5$ au) \citep{ZhuWu2018,fernandes}. A subset of these systems with inner sub-Neptunes have high obliquities (see Figure \ref{fig:psi_teff}). Though a range of formation models are still in play for close-in planets in general, the substantial gaseous envelopes of these planets indicate that they coexisted with a protoplanetary disk at some point in their evolution \citep{lee2016}. We describe below our motivation for a simplified physical model of a two planet system with an outer, slowly dispersing, protoplanetary disk. We also derive an analytic model for the secular evolution of such a system.

\subsection{Initial conditions}

The innermost regions of protoplanetary disks are complex environments whose properties are likely set by the interplay between high energy stellar radiation and magnetic fields \citep{dul10,pascucci}. The large and diverse population of ``transition" disks (those with inner regions depleted of gas, dust, or both) indicate that planetary systems interior to 1 AU might coexist with a more massive, external disk \citep[e.g.][]{Espaillat:2014,Andrews:2018}.
These observations motivate our simplified model in which the (dynamically relevant) protoplanetary disk lies exterior to the orbit of
any Jovian planet located at $\gtrsim1$ au.

We consider systems composed of two planets with
masses $M\i$ and $M\out$, evolving secularly in the presence of an outer gas disk. The disk is assumed to follow a Mestel profile ($M(<r) \propto r$), with a total mass $M_{\rm disk}(t)$,
and inner and outer radii given by $R\i$ and $R\out$, respectively.
In addition to the mutual perturbations between the planets, the outer 
planet is coupled to the gravitational potential of the disk, while
the inner planet is coupled to the quadrupolar field induced by stellar rotation and undergoes apsidal precession from post-Newtonian effects.
The planet orbital elements are $a\i$, $e\i$, $I\i$, $\omega\i$ and $\Omega\i$
for the inner planet, and similarly for the outer planet.

We evolve the system throughout the gas dispersal phase, which is short
enough for tidal dissipation with the star to be ignored. The system
is assumed to have formed in near-alignment (i.e., with small obliquities
and relative inclinations). Thus, any high inclinations 
are generated self-consistently, which is an important distinctive feature of this model. 

\subsection{Resonantly excited inclinations}
Inclinations can be resonantly excited 
if the nodal precession rates of the inner and outer planets
encounter a commensurability \citep[e.g.][]{war76}.
In the presence of an external disk, the nodal precession rate of
the outer planet is proportional to $M_{\rm disk}$ and typically fast ($|\dot{\Omega}\out|\gg|\dot{\Omega}\i|$). As the disk disperses, 
$|\dot{\Omega}\out|$ decreases, inevitably reaching  ($|\dot{\Omega}\out|\approx|\dot{\Omega}\i|$) in a process termed ``secular resonance passage'' or ``scanning secular resonances'' \citep{hep80,war81}.

The Hamiltonian of the secular system (Equation~\ref{eq:full_hamiltonian})
can be reduced to a simplified model for $e\i=0$ (Equation~\ref{eq:H_resonant}). The simplified model mimics  
the `second fundamental model of resonance' \citep{HL1983}, which is a one-degree-of-freedom
Hamiltonian with a pair of canonically conjugate variables, and a conserved quantity (Equation \ref{eq:Theta_prime}) proportional to
\ba
\label{eq:A_conserved}
\mathcal{A}\equiv M\i a\i^{1/2}(1-\cos I\i)+M\out a\out^{1/2}(1-\cos I\out)~.~~~
\ea
The model has one free parameter $\Delta$,
which defines a ``distance to resonance''(Appendix~\ref{app:resonance})
\ba
\label{eq:delta}
\Delta(t)= \frac{2}{3}\left[\frac{1+\eta _\star}{I_{\rm out,0}}\right]^{2/3}\left[1-\xi_{\rm disk}(t)
\right],
\ea
where $\xi_{\rm disk}$ measures the relative precession rates of the outer planet (driven by the disk) and the inner planet ($\simeq |\dot{\Omega}\out|/|\dot{\Omega}\i|$), and $\eta _\star$ the relative strength of the stellar quadrupole and the two-planet interactions. These are defined as follows,
\ba
\label{eq:xi_disk}
\xi_{\rm disk}(t)=\frac{a\out^{9/2}(1-e\out^2)^{3/2}}{a\i^{3/2}R\i^2R\out}\frac{M_{\rm disk}(t)}{(1+\eta _\star) M\out},
\ea
and \citep[e.g.][]{tre09}
\ba
\label{eq:eta_star}
\eta _\star&=&
\frac{2J_2M_\star}{M\out}\frac{R _\star^2a\out^3}{a\i^5}(1-e\out^2)^{3/2}.
\ea
In Equation~(\ref{eq:eta_star}),
$J_2$ is the star's second zonal harmonic, which can be related to the stellar rotation period 
 $P _\star$ by \citep{sterne}
\ba
J_2\simeq\frac{k_2}{3}\frac{4\pi^2}{P_\star^2}\frac{R _\star^3}{GM _\star},
\ea
where $k_2$ is the tidal Love number, which
is $\simeq0.2$ for the fully convective, pre-main sequence (PMS) stars that we consider here
(e.g., \citealt{claret}). 

Resonance crossing occurs
when $\Delta=0$, i.e. when $\xi_{\rm disk}=1$ (Eq.~\ref{eq:delta}). In this simplified model, resonant capture is {\it guaranteed} if the following conditions are met
\citep{HL1983}: (1) $\dot{\Delta}>0$ when $\Delta=0$, which requires a decaying disk with initially enough mass such that  $\xi_{\rm disk}>1$; (2) the starting inner planet inclination $I_{\rm in,0}$ is sufficiently low,
so $I_{\rm in,0}<I_{\rm in, cap}\propto [I\out /(1+\eta_\star)]^{1/3}$
(Equation~\ref{eq:capture_inclination});
and (3) the resonance is crossed with a  sufficiently small $\dot{M}_{\rm disk}$ to preserve adiabatic invariance (Equation~\ref{eq:disk_evolution_timescale}). 

The constraint that the resonance is crossed ``adiabatically"
can be written as
\ba
\label{eq:adiabatic_condition}
\tau_{\rm adia}< \left|\frac{d\log M_{\rm disk}}{dt}\right|^{-1}\equiv \tau_{\rm disk}(t)
\ea
where
\ba
\label{eq:t_adiabatic}
\tau_{\rm adia}\simeq \frac{2P\i}{3\pi}\frac{M _\star}{M\out}\frac{a\out^3}{a\i^3}
(1-e\out^2)^{3/2}
I_{\rm out,0}^{-4/3}\left(1+\eta _\star\right)^{1/3}~~~
\ea
is the adiabatic time, $P_{\rm in}$ is the inner planet's orbital period,
and $\tau_{\rm disk}$ is the disk dispersal time, which can itself be a function of time.
The degree of adiabaticity can be quantified in the ``adiabatic parameter''
$x_{\rm ad}\equiv \tau_{\rm disk}/\tau_{\rm adia}$.
As we show in Section~\ref{sec:predicted}, the three conditions for resonance capture are met for a wide range of realistic initial conditions.

During resonant capture, the system follows a slowly evolving fixed point in phase space,  which 
corresponds to
 $\Omega\i-\Omega\out=\pi$ and
\ba
\label{eq:I_analytic}
\cos I\i(t)=1- \frac{\left[x^*(t)\right]^2}{2}
\left[\frac{I_{\rm out,0}}{2(1+\eta _\star)}\right]^{2/3}
\ea
where
\begin{equation}
x^*=\left\{\!\!\!
\begin{array}{cc}
     \left(1+\sqrt{1-\Delta^3}\right)^\frac{1}{3} 
+\Delta\left(1+\sqrt{1-\Delta^3}\right)^{-\frac{1}{3}},
& \Delta\leq1\\\\
2\sqrt{\Delta}\cos\left(\tfrac{1}{3}\tan^{-1}\sqrt{\Delta^3-1}\right),
& \Delta>1
\end{array}
\right.
\end{equation}
\citep{PMT2012}. For $\Delta\gg1$, $x^*\approx \sqrt{3\Delta}$. Therefore, after the resonance
has been crossed, and $\xi_{\rm disk}\rightarrow0$, it is easy to check that
$\cos I\i\rightarrow0$, i.e. the inner orbit inexorably
approaches a polar configuration, if it remains circular. The latter constraint represents the aforementioned second phase of our mechanism, which we describe below.

\subsection{Exponential eccentricity growth and resonance detuning}
In the simplified treatment of resonant capture, we have assumed $e\i=0$ and arbitrary $e\out$ at quadrupolar order\footnote{We have checked numerically that octupole-level corrections play a minor dynamical role due to strong relativistic and $J_2$ precession, at least for $e\out\lesssim0.6$ in our fiducial set-up.}.

A simplified linear stability analysis of the inner orbit (Appendix~\ref{app:instability})
shows that initially circular orbits are unstable to eccentricity growth when
\ba
\label{eq:unstable_region}
\left(\frac{4+4\eta _\star+\eta_{\rm GR}}{10+5\eta _\star}\right)<\sin^2I\i<\left(\frac{4+4\eta _\star+\eta_{\rm GR}}{5\eta _\star}\right),
\ea
where 
\ba
\label{eq:eta_gr}
\eta_{\rm GR}=\frac{8GM _\star}{ c^2}
\frac{a_{\rm out}^3}{a\i^4}
\frac{M _\star}{M\out}
(1-e\out^2)^{3/2}
\ea
measures the relative strength of GR corrections with respect to the two-planet interaction.
For fiducial parameters, $\eta_{\rm GR}\sim20$, which inhibits eccentricity
growth  \citep{FT2007,liu15}.

 \begin{figure*}[t]
\center
\includegraphics[width=18.5cm]{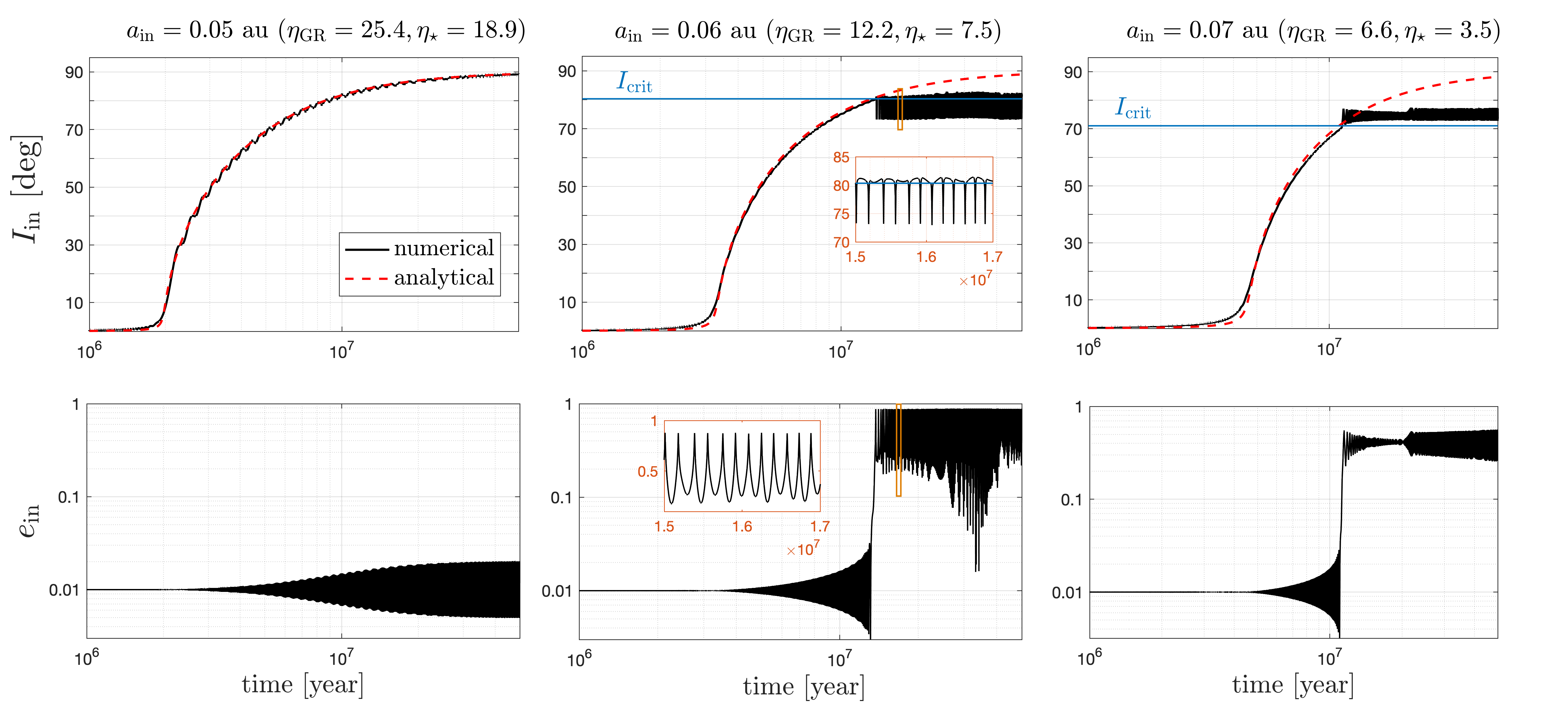}
\caption{Inclination and eccentricity evolution of a Neptune-mass planet orbiting a Solar-mass star with an initially nearly circular ($e\i=0.01$) and coplanar orbit ($I\i=1^\circ$ relative to the host star's equator). We place a 4 $M_J$ gas giant at 2 au in a circular orbit with inclination $I_{\rm out}=5^\circ$,  and a coplanar disk (relative to star's equator) with an inner edge at $3$ au whose mass  decays as $M_{\rm disk}=50M_J/[1+t/(1 {\rm Myr})]^{3/2}$.  The star has a radius of $1.3R_\odot$, a Love number $k_2=0.2$, and spin period of $P _\star=7$ days. In the {\bf left panels} we set $a\i=0.05$, satisfying the stability condition $\eta_{\rm GR}>6+\eta _\star$, thus leading to resonance capture into a polar orbit and no eccentricity instability. The red dashed line shows the analytical model from Equation (\ref{eq:I_analytic}) that perfectly reproduces the numerical integrations.  In the {\bf middle panels} we set $a\i=0.06$, predicting an instability at $I_{\rm crit}=81.3^\circ$ from Equation (\ref{eq:Icrit}), leading to exponential eccentricity growth up to $e\i\sim0.9$ and detuning of the resonance. The eccentricity-inclination oscillations are shown as a zoom-in inset in the orange boxes.  In the {\bf right panels}, we set $a\i=0.07$ resulting in $I_{\rm crit}\simeq71^\circ$ and eccentricity growth up to $\sim0.5$. The subsequent tidal evolution is ignored in this example as we focus mainly on the inclination excitation. }	
\label{fig:evol_example}
\end{figure*}

Because $I\i$ approaches the unstable region (Equation~\ref{eq:unstable_region})
from below, the relevant threshold is
\ba
\label{eq:Icrit}
I_{\rm crit}=\sin^{-1}\left( \frac{4+4\eta _\star+\eta_{\rm GR}}{10+5\eta _\star}\right)^{1/2}
\ea
which is a generalization of the well-known Lidov-Kozai critical angle $I\i\simeq39.2^\circ$,
recovered when $\eta_\star=\eta_{\rm GR}=0$. 

An important consequence from Equation (\ref{eq:Icrit}) is that all inclinations are stable if
\ba
\label{eq:stab_cond}
\eta_{\rm GR}\geq 6+\eta _\star 
\ea
in which case the resonant mechanism would pump inclinations all the way to $90^\circ$ while the orbit remains circular
(Equation~\ref{eq:I_analytic}).
In \citet{liu15}, the authors also consider the effect
of oblateness, but only for zero-obliquity, in which
case $J_2$ can only amount to a stabilizing effect.
Indeed, from equation 50 of that paper, one can derive that
the unconditional stability requirement in such a case is $\eta_{\rm GR}>6-\tfrac{4}{3}\eta _\star$.
Both conditions reduce to Equation (36) of \citet{FT2007} when $\eta_\star=0$.

The limit of $\eta_\star\gg1$ and $\eta_{\rm GR}\approx0$ is also interesting. In this
case, $I_{\rm crit}\approx63.4^\circ$, known as the ``critical inclination'' in geo-satellite dynamics, 
which marks the boundary between prograde to retrograde apsidal precession.
Around $63.4^\circ$, there is a narrow unstable region of width $\Delta I=2/\eta _\star$.
Therefore, in this limit, resonance detuning takes place at $I\i\approx63^\circ$,
saturating the final inclination to this value. Conversely, for $I_{\rm crit}$ to be greater than $63.4^\circ$, one must
require
\ba
\label{eq:polar_cond}
6+\eta _\star>\eta_{\rm GR}>4
\;\;\;\;\text{(eccentric, inclined orbits)}~~~
\ea
Consequently, values of
$\eta_{\rm GR}$ greater than 4 are instrumental in overcoming this early-onset saturation 
of inclination, and in tilting orbits toward nearly polar configuration. In this sense, the creation of polar-orbit planets is inherently a post-Newtonian effect.

\section{Predicted obliquities}\label{sec:predicted}

\subsection{Behavior of the fiducial system}
To test the predictions of the analytical model, 
we numerically integrate the full equations of motion (\ref{eq:Milanko_1}-\ref{eq:Milanko_3})
for a range of parameters and initial conditions. The parameter
space may appear hopelessly multi-dimensional, but most
of the physics is contained in the values
 of $\eta_{\rm GR}$ and $\eta_\star$, which determine
if and when inclination growth is 
saturated via resonance detuning.

In Figure \ref{fig:evol_example}, we show two examples
of an initially coplanar Neptune-mass planet that undergoes inclination growth, with
 $\eta_{\rm GR}\simeq25.3$ and $\eta _\star\simeq18.9$ (left panels),
and with
$\eta_{\rm GR}\simeq12.2$ and $\eta _\star\simeq7.6$ (right panels).
In the first case, condition (\ref{eq:stab_cond}) is satisfied, and the orbit
reaches a final inclination of $90^\circ$ (black line, top)
while remaining nearly circular ($e\i \lesssim.02$) (black line, bottom).
In the second case, only condition (\ref{eq:polar_cond}) is satisfied, and the inclination grows to $I_{\rm crit}\simeq81.3^\circ$ (black line, top), as predicted by Equation~(\ref{eq:Icrit}). As $I_{\rm crit}$ is reached, eccentricity grows exponentially
until quasi-regular eccentricity-inclination oscillations are established (see the zoom-in inset in middle panels). We overlay in red the theoretical (adiabatic) inclination growth given by 
Equation~(\ref{eq:I_analytic}). In both examples, the agreement is excellent.

 \begin{figure*}[t]
\center
\includegraphics[width=\textwidth]{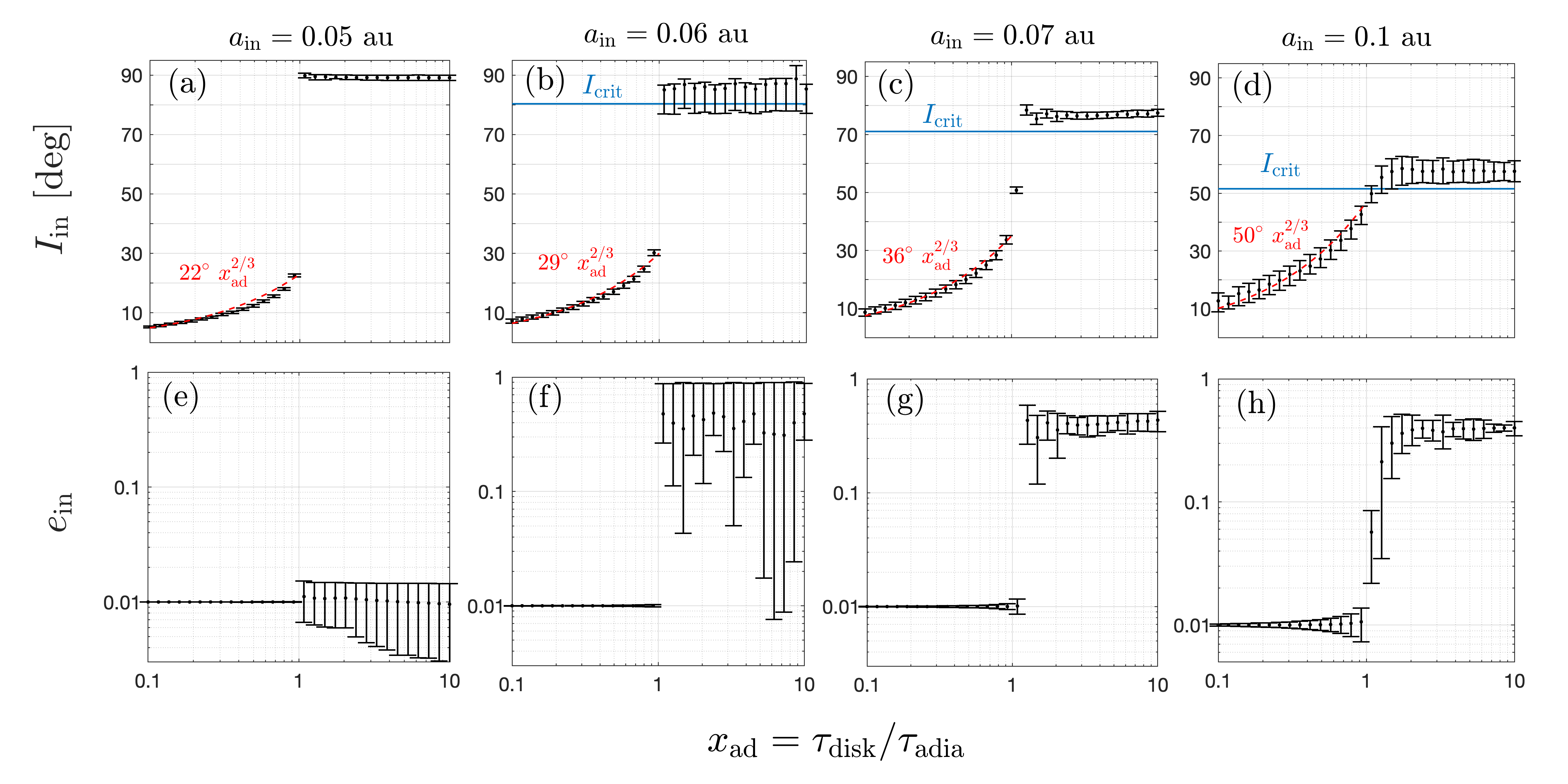}
\label{fig:adiabatic}
\caption{Post-resonance inclinations and eccentricities as a function of the disk depletion timescales  expressed as a function of the adiabaticity parameter $x_{\rm ad}=\tau_{\rm disk}/\tau_{\rm adia}$  for Neptune-like planets at 0.05 au (panels a and e), 0.06 au (b and f), 0.07 au (c and g), and 0.06 au (d and h). The other parameters are the same as in figure \ref{fig:evol_example}, except that the disk is assumed to decay exponentially so $\tau_{\rm disk}=d\log M_{\rm disk}/dt$ is constant in time. The error bars indicate the minimum and maximum values centered at the mean calculated over a window of time in $[9\tau_{\rm disk},10\tau_{\rm disk}]$. All panels show the transition from a non-adiabatic resonance crossing at $x_{\rm ad}<1$  to an adiabatic one above $x_{\rm ad}>1$. The former leaves the eccentricities  unperturbed and excites only moderate inclinations increasing with  $x_{\rm ad}$ as $x_{\rm ad}^{2/3}$ (see fitted lines). In turn, the adiabatic cases reach final inclinations in agreement with our predicted values, where for stable (GR-dominated) systems reach inclinations of $90^\circ$ (panel a), while the unstable cases reach values close to $I_{\rm crit}$ (Eq. [\ref{eq:Icrit}], shown in horizontal blue lines). In the unstable cases, the final eccentricities reach order unity, undergoing large-amplitude  $e\i-I\i$ oscillations.}
\end{figure*}

\subsection{Numerical experiments: assessing the adiabaticity}

In Figure \ref{fig:adiabatic} we show the values of the inner planets inclination and eccentricity long after the resonance is crossed from a suite of numerical experiments where we vary the disk dispersal timescale given in units of the adiabaticity parameter $x_{\rm ad}=\tau_{\rm disk}/\tau_{\rm adia}$. Each panel from left to right corresponds to a different semi-major axis $a\i$ and the other parameters are the same as in Figure~\ref{fig:evol_example}. We observe that whenever a system evolves adiabatically, i.e., when $x_{\rm ad}>1$, there is resonant capture (inclination grows toward $I_{\rm crit}$), in accordance with the theory. 
On the other hand, for non-adiabatic resonance passage, the planet still receives a kick in inclination, $I_{\rm non-ad}$ \citep[e.g.,][]{quillen2006}. The magnitude of this excitation is empirically well described by
\ba
\label{eq:I_nonad}
I_{\rm non-ad}\simeq 22^\circ \left[\frac{I_{\rm out,0}}{4^\circ}\cdot\frac{20}{(1+\eta _\star)} \right]^{1/3} x_{\rm ad}^{2/3}
\ea
(red lines in Figure~\ref{fig:adiabatic}). In most cases, $I_{\rm non-ad}<I_{\rm crit}$, 
which means that the eccentricity instability
is not triggered, and the orbits remain circular. 

All the systems captured into resonance have post-capture inclinations that are consistent with either the predicted polar state for stable systems (panel a with $a\i=0.05$ au), or with $I_{\rm crit}$ for the unstable systems (panels b, c, and d).  The post-capture eccentricities of the unstable systems (panels f, g, h) oscillate in time. Conversely, systems that are {\it not captured} into resonance (with adiabaticity parameter $x_{\rm ad}<1$) exhibit moderate inclination growth with ($I\sim10-40^\circ$) and no eccentricity excitation.

\subsection{Population predictions}

Having established the final orbital states long after the disk dispersal, we can make predictions for the final stellar obliquities  as a function of disk properties ($t_V$ and $M_{\rm disk,0}$), stellar properties ($P _\star, R _\star$), planetary architecture ($a\i,M\out,a\out$) and the initial inclination of the outer planet $I_{\rm out,0}$.

Our procedure to obtain the final inclination $I_{\rm final}$ is as follows.
\begin{enumerate}
\item We determine if $\xi_{\rm disk}(t=0)>1$ (Eq. [\ref{eq:xi_disk}]) and  the resonance is crossed . If the resonance is not crossed, then $I_{\rm final}=0$.
\item We assess the adiabaticity of the resonance crossing. If  $x_{\rm ad}>1$ (adiabatic), then $I_{\rm final}=I_{\rm crit}$. If $x_{\rm ad}<1$ (non-adiabatic), then $I_{\rm final}=I_{\rm non-ad }$ from Equation (\ref{eq:I_nonad})
\end{enumerate}

\begin{figure*}[t!]
\center
\includegraphics[width=18.8cm]{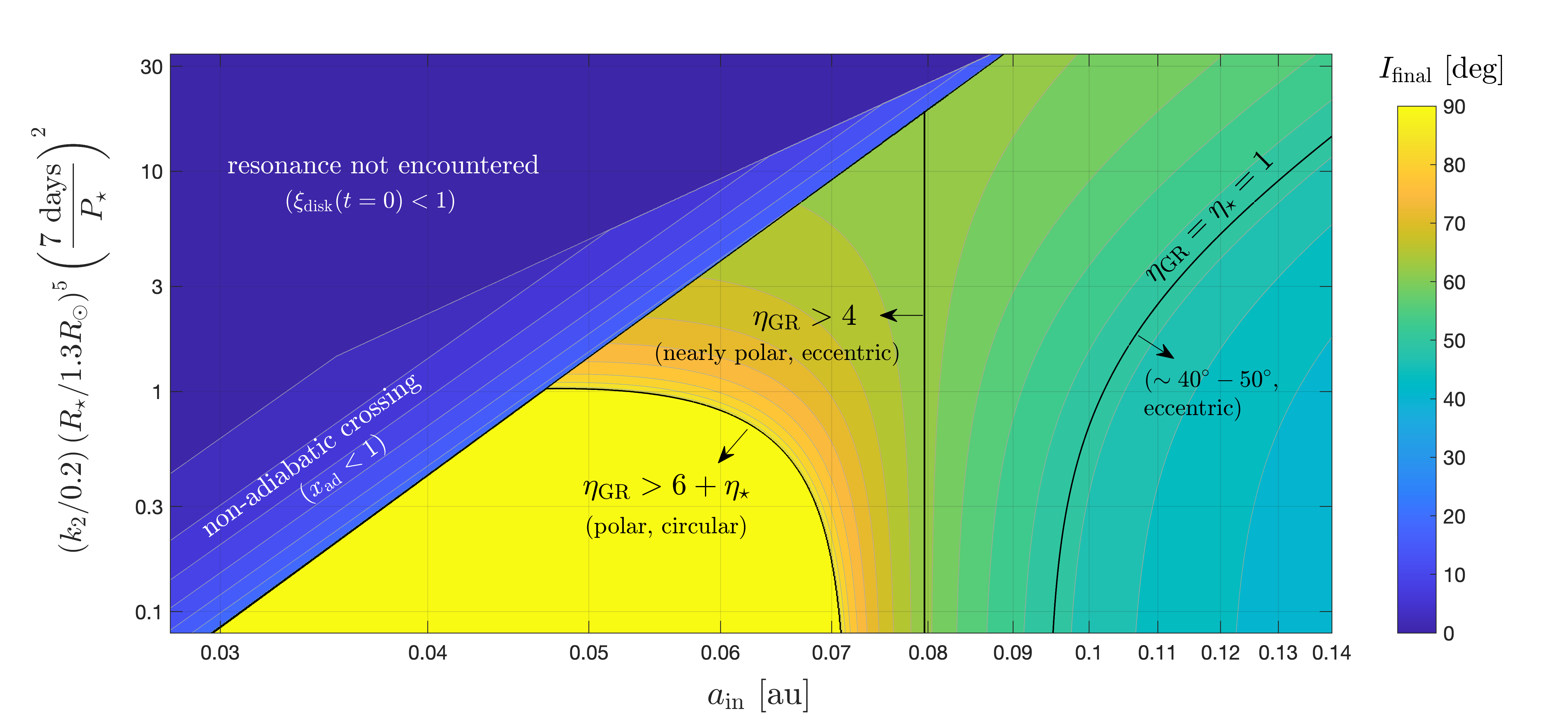}
\vspace{-0.1in}
\label{fig:contour}
\caption{Final stellar obliquities as a function the semi-major axis of the inner planet $a\i$ and the rotationally-induced stellar oblateness represented by the combination $k_2R _\star^5/P _\star^2$.  We fix the outer planet properties ($M\out=4M_J$, $a\out=2$ au, and $I\out=5^\circ$) and disk evolution  as $M_{\rm disk}=50M_J/(1+t/1{\rm Myr})^{3/2}$. Large obliquities are attained in the region where the resonance is crossed ($\xi_{\rm disk}(t=0)>1$ in Eq. \ref{eq:xi_disk}) and the crossing is adiabatic ($x_{\rm ad}= \tau_{\rm disk}/\tau_{\rm adia}>1$ in Eq. \ref{eq:adiabatic_condition}). Within this region, the planets acquire nearly polar orbits for $\eta_{\rm GR}>4$ at $a\i \lesssim0.08$ au, and eccentricity excitation occurs when $\eta_{\rm GR}<6+\eta _\star$. The lower-right region is dominated by the outer planet ($\eta_{\rm GR}, \eta _\star<1$) and reaches obliquities of $\lesssim 50^\circ$.}	
\end{figure*}

In Figure \ref{fig:contour}, we show the final inclination $I_{\rm final}$ as a function of $a\i$ and the stellar properties that determine the $J_2$ potential $k_2R _\star^5/P _\star^2$. The resonance is only encountered outside the blue region where the stellar quadrupole is weak enough. Here, we identify two distinct regions in parameter space:
\begin{enumerate}
    \item a region dominated by relativistic precession with $\eta_{\rm GR}>4$ that leads to nearly polar orbits at $a\i\lesssim 0.08$ au (yellow to orange countours), including a region that is stable to eccentricity perturbations at $\eta_{\rm GR}>6+\eta _\star$;
    \item a region where the precession is dominated by the outer planet with $a\i\gtrsim 0.1$ au and $\eta _\star,\eta_{\rm GR}<1$  reaching inclinations of $\sim40^\circ-50^\circ$ ($I_{\rm crit}<51.7^\circ$).
\end{enumerate}

\section{Application to observed systems} 
\label{sec:observed}

For any known close-in Neptune in a tilted orbit,
we can use the above procedure to predict
the orbital properties of an outer companion. As a proof of concept, we focus on the HAT-P-11 system, where the nearly polar inner planet has a known outer companion HAT-P-11c \citep{yee}. Given the semi-major axis of HAT-P-11b (0.052 au) and reasonable assumptions for the disk dispersal time, and for the PMS stellar radius and rotational period, the resulting obliquity becomes a function of only $M\out$ and ${b\out}\equiv a\out(1-e\out^2)^{1/2}$, the unseen companion's mass, and its semi-minor axis, respectively.

\begin{figure*}[t!]
\center
\includegraphics[width=18.2cm]{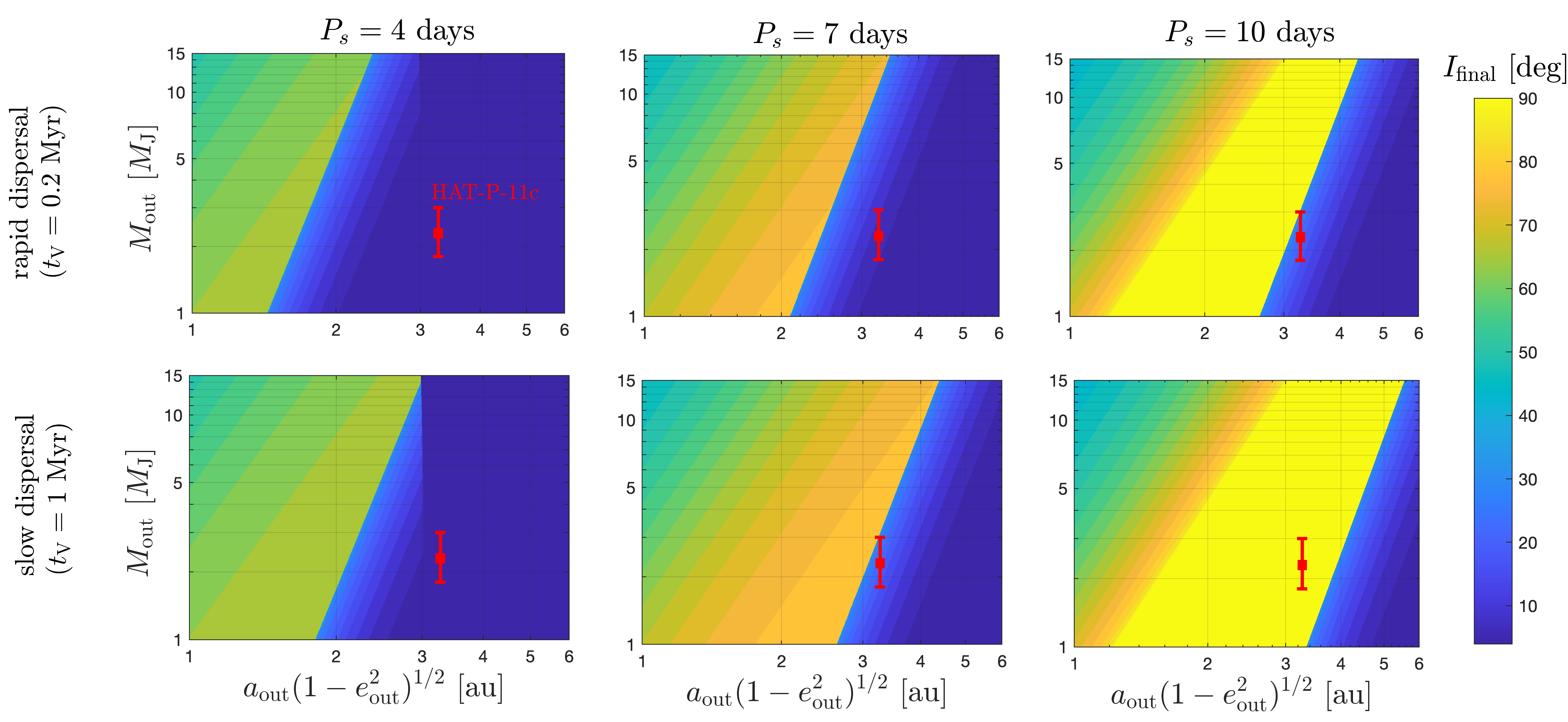}
\label{fig:contour_hat}
\caption{Final obliquity as a function of the outer planet's mass $M\out$ and semi-minor axis $a\out(1-e\out^2)^{1/2}$ for different rotations periods of the hosts $P_s$ (4, 7, and 10 days from left to right) and two disk models with $t_V=0.2$ Myr (rapid dispersal, upper panels) and $t_V=1$ Myr (slow dispersal, lower panels) with $M_{\rm disk}=50M_J/(1+t/t_V)^{1/2}$.  The host star has a mass of $M_\star=0.8M_\odot$ similar to HAT-P-11, the planet's semi-major axis at $0.052$ au and we set its radius to $1.3R_\odot$ (typical of K-dwarfs with ages of several Myrs, \citealt{baraffe}). The error bar indicates the measurement for HAT-P-11c \citep{pimen1}. Note that its current radius and rotation period are $R _\star\simeq 0.68R_\odot$  $P _\star\simeq29$ days \citep{yee}.
}	
\end{figure*}

In Figure \ref{fig:contour_hat}, we show the expected obliquity as a function of $M\out$ and $b\out=a\out(1-e\out^2)^{1/2}$ assuming various rotation periods representative of low-mass PMS stars \citep{bouvier2013},
and for rapid and slow dispersal (top and bottom panels, respectively).
From the figure, we see that polar orbits (orange-to-yellow regions) are produced with great likelihood if $P _\star=10{\rm d}$ (right panels)
and to a moderate extent  $P _\star=7{\rm d}$ (middle panels).
The known values for HAT-P-11c are included in each panel (red squares), with
a predicted ``high obliquity'' region in the rightmost panels.

In conclusion, provided that the star rotates slowly enough and the disk is sufficiently long-lived (typically $\sim3$ Myr), our model can explain the large obliquity of HAT-P-11b and possibly the low stellar obliquity for the outer planet as the mutual inclination is consistent with $\sim 90^\circ$ ($54^\circ<i_{\rm b,c}<126^\circ$ at 1-$\sigma$; \citealt{pimen1}). The nearly polar state is expected as $\eta_{\rm GR}\simeq 55$, much larger than the required threshold of 4 (Figure \ref{fig:contour}).

\subsection{Other tilted systems}

We can extend the analysis for HAT-P-11 to other tilted systems based on their current orbital states, noting that nearly polar planets should reside in systems with $\eta_{\rm GR}>4$, while those with moderate obliquities ($\lesssim 50^\circ$)  $\eta_{\rm GR}<4$ (or a non-adiabatic crossing).  Using these constraints, we both confirm the viability of our mechanism for systems with known cold Jovians, and predict the properties of the planets yet to be detected. We exclude the compact multi HD-3167 and 
Cancri-55 shown in Figure \ref{fig:psi_teff}, see Section \ref{sec:future}:
\begin{itemize}
    \item {\it $\pi$ Mensae} has an obliquity of $\simeq {27^\circ}^{+5.8^\circ}_{-4.7^\circ}$ \citep{triaud2020}, $M_\star\simeq 1.1 M_\odot$, $a\i=0.068$, $b\out=2.54$ AU, and $M\out\simeq 14M_{\rm J}$, leading to $\eta_{\rm GR}\simeq 1.3$, consistent with the non-polar orbit expectation (provided an adiabatic crossing). Also, mutual inclination between b and c is $49^\circ<i_{\rm b,c}<131^\circ$ at 1-$\sigma$ barely consistent with a low-obliquity Jovian, but consistent at 2-$\sigma$ \citep{pimen1};
    \item {\it WASP-107} has a near polar orbit, while $a\i\simeq0.055$ AU and $M_\star\simeq0.69M_\odot$, thus requiring $(b\out/2\mbox{ AU})^3\gtrsim(M\out/0.5M_J)$;
    \item {\it GJ-436} also has a nearly polar orbit, while $a\i=0.28$ AU and $M_\star\simeq 0.4 M_\odot$, thus requiring a companion with $(b\out/3\mbox{ AU})^3\gtrsim(M\out/1.2M_J)$.
    \item {\it Kepler-408} has an obliquity of ${48^\circ}^{+4^\circ}_{-5^\circ}$, while $a\i\simeq0.037$ AU and $M_\star\simeq1.05M_\odot$, compatible with either a non-adiabatic resonance passage or a capture with $\eta_{\rm GR}\lesssim 1$ (i.e, $[b\out/0.28\mbox{ AU}]^3\lesssim [M\out/1M_J]$);
\end{itemize}
The detection of Jovian-mass companions with the predicted properties will provide strong support to our model as well as the measurements of low obliquities of cold Jupiter systems, the first of which measurements was performed using interferometry  in the $\beta$ Pictoris system, finding strong evidence for low obliquities \citet{betapic}.

\section{Discussion}

For the first time, we have analytically demonstrated that
a nearly co-planar system of two planets and a disk can secularly evolve into one with high obliquities and eccentricities (for the inner planet) and large mutual inclinations (with the still co-planar outer Jovian).

The novelty of this mechanism is that it can self-consistently produce close-in planets that are highly inclined {\it and} eccentric \citep[see][]{correia}, without invoking extreme initial conditions, i.e. large primordial misalignments of stellar equators, disks, planets or some combination therein. Instead it relies on the natural dissipation of the protoplanetary disk to induce resonance sweeping and capture.

This discovery required substantial developments beyond the classic Lagrange-Laplace theory \citep{hep80,war81}.  We have worked out a proper non-linear resonance, valid for arbitrary inclinations, and for which resonant``capture'' is well defined. The mathematical formalism of this treatment is closely
related to that of \citet{bat16}, where the authors attributed resonance sweeping to a decline in stellar oblateness \citep{war76}, rather than disk mass\footnote{In our set-up, a waning stellar quadrupole  does not lead to capture as the resonance is crossed  from the wrong direction ($\dot{\Delta}<0$ in Eq. \ref{eq:delta}). However, the set-up in  \citet{bat16}, where the test particle is outside of the Jovian does cross from right direction.}.

 Finally, this mechanism makes specific predictions for the required properties of as-yet undetected outer planets that should be easily testable with ongoing radial velocity surveys and astrometric measurements from Gaia.

\subsection{Caveats and future work}
\label{sec:future}
While we have shown that many of the polar planets in Figure \ref{fig:psi_teff} are easily produced by our model, we highlight several areas that require future study.

\paragraph{Are the orbital configurations sustained on Gyr timescales?} We have thus far carried out integrations of the systems for up to $\sim 10$ Myrs. The most likely culprit to alter orbits on Gyr timescales is the tidal dissipation of the residual eccentricities, also damping the planet's semi-major axis. This orbit shrinkage would act to further decouple the sub-Neptune from the outer planet due to enhanced relativistic precession, effectively freezing the inclinations at their large values, not altering our results.  

\paragraph{Can compact multi-planet systems be resonantly tilted?} The resonant excitation of inclinations could readily operate in a compact multi-planet system, but the danger lies in the eccentricity instability at high inclinations which can lead to close encounters and destabilization of the close-in planets. However, similar to the role of general relativistic precession, the planet-planet interactions may act to stabilize the system against the eccentricity instability \citep{denham}. As such, our model could provide a sound mechanism to account for systems such as Kepler-56 \citep{huber2013} and the polar multi-planet system HD-3167 \citep{dalal}. 

\paragraph{Does the resonance affect hot Jupiter systems?} While there is no upper mass limit for excitation, the larger masses of hot Jupiters compared to sub-Neptunes would demand  initial inclinations for the outer planet that are larger by a factor of $\sim 3-10$ to satisfy the conservation of  conservation of  angular momentum deficit (Equation \ref{eq:A_conserved}). Specifically, the inner planet can reach a polar orbit only for $I\out\gtrsim (m\i/m\out)^{1/2}(a\i/a\out)^{1/4}$ leading to $I\out\gtrsim 3.7^\circ$ in our fiducial Neptune and $I\out\gtrsim 11.7^\circ$ for a hot Jupiter. Because the mechanism no longer operates in the nearly co-planar limit, we deem it less promising, though similar conditions are invoked in other models for high obliquity hot Jupiters \citep{matsakos}.

\paragraph{How does the stellar type affect the resonance?} Inclination excitation is most likely when the rotationally-induced stellar quadrupole is small, and disks longer-lived. The former condition promotes  resonant capture, while the latter promotes the adiabaticity of the resonant encounter. These two constraints operate in tandem to favor lower-mass stars. First, they are naturally smaller in radius, even  with their slower pre-main sequence contraction \citep{baraffe}. Secondly, low mass stars  harbor longer-lived disks \citep{disk_life}. Finally, resonance crossing occurs at later times, and thus smaller $R_*$, for slowly dissipating disks. This preference appears to be borne out observationally: polar planetary systems are hosted by M to K dwarfs (see Figure \ref{fig:psi_teff}).

\section{Conclusions}

We have proposed a novel mechanism to explain the orbital architectures of a population of sub-Neptunes in non-circular, nearly-polar orbits (stellar obliquities of ${\sim}90^\circ$) with misaligned outer companions.

The mechanism consists of a joint process of resonance sweeping and parametric instability, driven by disk dispersal. 
A long enough dispersal timescale guarantees resonant capture and subsequent inclination growth. The inclination growth is then halted by the eccentricity instability threshold, in turn leading to eccentricity growth. The inclination threshold is pushed to large values primarily by post-Newtonian corrections, making General Relativity a fundamental factor in producing polar orbits.

This mechanism predicts that nearly polar sub-Neptunes should coexist with cold Jupiters in low stellar obliquity orbits and orbital periods that are long enough so that the planet's apsidal precession is dominated by relativistic effects ($\eta_{\rm GR}>4$).


\acknowledgements 
We are grateful to Andrew Youdin, Chris Spalding, Dan Tamayo, Eric Ford, Fei Dai, Josh Winn, J.J. Zanazzi, Kento Masuda, Max Moe, Ryan Rubernzahl, Sarah Millholland, Yubo Su, and Wei Zhu, for stimulating and useful discussions.
CP acknowledges support from the Bart J. Bok fellowship at Steward Observatory. DJM acknowledges support from the CIERA Fellowship at Northwestern University.
RM is grateful for research support from NSF (grant
AST-1824869), NASA (grant 80NSSC18K0397) and the Louise
Foucar Marshall Foundation. 

\bibliography{refs}

\appendix

\section{A. Equations of motion and definitions}
\label{app:equations_of_motion}

It is easiest to express the potential that includes the secular coupling between the planets and the external fields due the stellar quadrupole (oriented along ${\bf \hat{s}}$) and the disk (oriented along ${\bf \hat{j}}_{\rm disk}$) in terms of the eccentricity vectors ${\bf e}=e{\bf \hat{e}}$ and specific angular momentum vectors ${\bf j}=(1-e^2)^{1/2}{\bf \hat{j}}$. By defining the indices 'in' and 'out' the vectors (and orbital elements later on) for the inner and outer planets, the potential reads (e.g., \citealt{TY2014}):
\ba
\label{eq:full_hamiltonian}
\phi &=&  
-\frac{\phi_{\rm in,\star}}{2} \left[ 
\frac{ ({\bf \hat{s}}\cdot \jin)^2-\tfrac{1}{3}j\i^2}{j\i^5}
\right]-
\frac{\phi_{\rm in, GR}}{2j\i}
-\frac{\phi_{\rm in,out}}{2}\left[-5( \ein \cdot \jout )^2 +(\jin \cdot \jout  )^2
+2e\i^2-\tfrac{1}{3}\right]
-\frac{\phi_{\rm out, disk}}{2}({\bf \hat{j}}_{\rm disk}\cdot \jout)^2
\nonumber\\
\ea
where the amplitudes are
\ba
\phi_{\rm in,\star}&=&\frac{3J_2GM\i M _\star R _\star^2}{2a\i^3},\\
\phi_{\rm in, GR}&=&\frac{6G^2M\i M _\star^2}{a\i^2 c^2},\\
\phi_{\rm in, out}&=&\frac{3GM\i M\out a\i^2}{4b^3\out},
\ea
with $b\out=a\out(1-e^2\out)^{1/2}$ the semi-minor axis of the outer planet. We note that writing the equations of motion in terms of orbital elements is cumbersome, and decided to evolve the full system using vectors, while carrying out the analytic calculations in Appendices \ref{app:resonance} and \ref{app:instability} using orbital elements for limiting cases.

For the disk, we model its potential using the distant tide approximation as in \citet{Terquem2010}, which for a Mestel disk with mass $M_{\rm disk}$ and inner and outer edges $R\i$ and $R\out$, respectively, results in
\ba
\label{eq:phi_mestel}
\phi_{\rm out,disk}&=&\frac{3G M\out M_{\rm disk}a\out^2(R\out+R_{\rm in})}{8R_{\rm in}^2R\out^2}\mathcal{B}\left(\frac{a\out}{R_{\rm in}}\right),
\ea
where we have included a multiplicative factor  $\mathcal{B}\left(a\out/R_{\rm in}\right)$ to correct the expression for the parts of the disk close to the planet as in \citet{petro2019}. We set $\mathcal{B}\left(a\out/R_{\rm in}\right)=2$ , valid for $R_{\rm in}/a\out\sim 1.5$, thus approximating the amplitude of the potential to
\ba
\phi_{\rm out,disk}&\simeq&\frac{3G M\out M_{\rm disk}a\out^2}{4R_{\rm in}^2R\out}.
\ea

We solve the motion of ${\bf e}\i, {\bf j}\i, {\bf j}\out$ using the Milankovitch set of equations (e.g., \citealt{TY2014}) as
\ba
\label{eq:Milanko_1}
\frac{d{\bf j}\i}{dt}&=& -\frac{1}{L\i}\left(\nabla_{{\bf j}\i } \phi \times \jin+
\nabla_{\ein }\phi \times \ein\right)
\\
\label{eq:Milanko_2}
\frac{d\ein}{dt}&=&-\frac{1}{L\i}\left(\nabla_{\ein }\phi \times \jin+
\nabla_{\jin }\phi \times \ein\right)
\\
\label{eq:Milanko_3}
\frac{d\jout}{dt}&=&-\frac{1}{L\out}\nabla_{\jout }\phi \times \jout,
\ea
where $L\i=M\i\sqrt{GM _\star a\i}$ and $L\out=M\out\sqrt{GM _\star a\out}$ are the angular momenta.

\section{B. Inclination resonance:  analytic model and conditions for capture}
\label{app:resonance}

We simplify the potential assuming that $e\i=0$ during the inclination resonance phase and write
\ba
\phi = -\tfrac{1}{2}\phi_{\rm in,out} ( \jout\cdot \jin )^2
-\tfrac{1}{2}\phi_{\rm in,\star}
({\bf \hat{s}}\cdot \jin)^2
-\tfrac{1}{2}\phi_{\rm out, disk} ({\bf \hat{j}}_{\rm disk}\cdot \jout)^2.
\ea
We express this potential as a two-degree-of-freedom Hamiltonian using  orbital elements defined relative to ${\bf \hat{s}}$ ($={\bf \hat{j}}_{\rm disk}$) as 
\ba
\mathcal{H}&=&-\tfrac{1}{2}\phi_{\rm in,\star}\cos^2 I\i
-\tfrac{1}{2}\phi_{\rm in,out}\big[\cos^2I\i\cos^2I\out+\tfrac{1}{2}\sin 2I\i\sin2I\out\cos \left(\Omega\out-\Omega\i\right)\nonumber\\
&&+\sin^2I\i\sin^2I\out\cos^2 \left(\Omega\out-\Omega\i\right)\big]
-\tfrac{1}{2}\phi_{\rm out, disk}\cos^2 I\out.
\ea
We express this Hamiltonian in Poincar\'e variables $\{-\Omega\i,Z\i=L\i(1-\cos I\i)\}$ and $\{-\Omega\out,Z\out=L\out(1-\cos I\out)\}$ approximating $\sin 2I\i\simeq 2\sqrt{2Z\i/L\i}$ and $\sin 2I\out\simeq 2\sqrt{2Z\out/L\out}$ and retaining only the lowest-order terms in $Z\out$. Thus,
\ba
\mathcal{H}\simeq-(\phi_{\rm in,out}+\phi_{\rm in,\star})\frac{(L\i-Z\i)^2}{2L\i^2}-
\phi_{\rm in,out}\sqrt{\frac{2Z\i}{L\i}}\sqrt{\frac{2Z\out}{L\out}}\cos \left[\Omega\out-\Omega\i\right]+(\phi_{\rm in,out}+\phi_{\rm out, disk})\frac{Z\out}{L\out}.
\ea
We perform a canonical transformation to the new  pairs $\{\theta,\Theta\}$ and $\{\theta',\Theta'\}$ using the following the generating function
\ba
\mathcal{F}=\left[\Omega\out-\Omega\i\right]\Theta-\Omega\out\Theta',
\ea
such that $\theta=d\mathcal{F}/d\Theta=\Omega\out-\Omega\i$, $Z\i=-d\mathcal{F}/d\Omega\i=\Theta$ and $Z\out=-d\mathcal{F}/d\Omega\out=\Theta'-\Theta$, and
\ba
\mathcal{H}\simeq-(\phi_{\rm in,out}+\phi_{\rm in,\star})\frac{(\Theta-L\i)^2}{2L\i^2}-
\phi_{\rm in,out}\sqrt{\frac{2\Theta}{L\i}}\sqrt{\frac{2(\Theta'-\Theta)}{L\out}}\cos\theta +(\phi_{\rm in,out}+\phi_{\rm out, disk})\frac{(\Theta'-\Theta)}{L\out}.
\ea
We note that the Hamiltonian does not depend on $\theta'$, implying that
\ba
\label{eq:Theta_prime}
\Theta'=L\i(1-\cos I\i)+L\out(1-\cos I\out)
\ea
is a constant of motion, stating that the angular momentum deficit is conserved. By dropping inessential constants and using that $L\i\ll L\out$ such that $\Theta\ll\Theta'$, we reduce the Hamiltonian to
\ba
\mathcal{H}\simeq
\left[\frac{(\phi_{\rm in,out}+\phi_{\rm in,\star})}{L\i}
-\frac{(\phi_{\rm in,out}+\phi_{\rm out, disk})}{L\out}\right]\Theta
-\frac{(\phi_{\rm in,out}+\phi_{\rm in,\star})}{2L\i^2}\Theta^2-
\phi_{\rm in,out}\sqrt{\frac{2\Theta}{L\i}}\sqrt{\frac{2\Theta'}{L\out}}\cos\theta.
\ea
Furthermore, assuming that inclinations are initially small, we can write $\Theta'\simeq L\out I_{\rm out,0}^2/2$. Similarly, it is safe to assume that $\phi_{\rm in,out}\ll \phi_{\rm out,disk}$, thus further simplifying the Hamiltonian
\ba
\mathcal{H}\simeq
\left[\frac{(\phi_{\rm in,out}+\phi_{\rm in,\star})}{L\i}
-\frac{\phi_{\rm out, disk}}{L\out}\right]\Theta
-\frac{(\phi_{\rm in,out}+\phi_{\rm in,\star})}{2L\i^2}\Theta^2-
\phi_{\rm in,out}I_{\rm out,0}\sqrt{\frac{2\Theta}{L\i}}\cos\theta.
\ea
Following \citet{HL1983} we can further simplify this Hamiltonian by re-scaling the variables as
\ba
\tau&=&\left(\frac{1+\eta _\star}{8}\right)^{1/3}\left( I_{\rm out,0} \right)^{2/3}\frac{t}{\tau_{\rm sec}} ,\\
\label{eq:R}
R&=& \left(\frac{1+\eta _\star}{I_{\rm out,0}}\right)^{2/3}\frac{\Theta}{L\i}=\left(\frac{1+\eta _\star}{I_{\rm out,0}}\right)^{2/3}(1-\cos I\i)\\
r&=& \pi-\theta=\pi-\Omega\out+\Omega\i,
\ea
with $\tau_{\rm sec}=L\i/\phi_{\rm in,out}$ and $\eta _\star=\phi_{\rm in,out}/\phi_{\rm in,\star}$, to arrive to the `second fundamental model of resonance':
\ba
\label{eq:H_resonant}
\mathcal{K}&=&-3\Delta R+R^2-2\sqrt{2R}\cos (r),
\ea
where 
\ba
\label{eq:Delta_appendix}
\Delta(t)&=& \frac{2}{3}\left[\frac{1+\eta _\star}{I_{\rm out,0}}\right]^{2/3}\left[1-\frac{\phi_{\rm out,disk}(t)\tau_{\rm sec}}{L\out(1+\eta _\star)}
\right]
\ea

 As shown by \citet{HL1983}, capture into resonance is certain if the following conditions are satisfied:
\begin{enumerate}
\item $d\Delta/d\tau>0$ as it crosses 0. This requires that initially the precession rate of the outer planet driven by the disk $\dot{\Omega}\out\simeq-\phi_{\rm out,disk}/L\out$ dominates over the precession rate of the inner planet driven by both the outer planet and the stellar rotationally-induced quadrupole given by $\dot{\Omega}\i\simeq-(1+\eta _\star)/\tau_{\rm sec}$ at $I\i\ll1$.
\item the action (i.e., the inclination) is small far from the resonance. More precisely that  $R_0<3$, or replacing Equation (\ref{eq:R}) with $Z_0/L\i\sim I_{\rm in,0}^2/2$, the initial inclination is
\ba
\label{eq:capture_inclination}
I_{\rm in,0}<3\left[\frac{I_{\rm out,0}}{1+\eta _\star}\right]^{1/3}.
\ea
The capture probability decays with $R_0>3$ \citep{HL1983}. In our applications $R_0<3$ always. 
\item $\Delta$ changes slowly near the resonance crossing. In particular, when $R_0\ll1$ we require that $d\Delta/d\tau'<g$ with $g$ of order unity, implying
\ba
\frac{d\Delta}{d\tau}=\frac{4}{3}~\tau_{\rm sec}^2\frac{|\dot{\phi}_{\rm out,disk}|}{L\out}
\left[\frac{1}{I_{\rm out,0}}\right]^{4/3}\left[\frac{1}{1+\eta _\star}\right]^{2/3}<g.
\ea
We numerically found that $g=4/3$ provides with a good threshold to capture into resonance up to a nearly polar orbit\footnote{Others numerical estimates for capturing planet into first-order mean-motion resonances yield a slightly larger value of $g\sim 2$ \citep{Friedland2001,quillen2006}} (see Figure \ref{fig:adiabatic} showing a mumerical test of adiabaticity). Since $\phi_{\rm out,disk}\propto M_{\rm disk}(t)$, the condition can be expressed in terms of the disk's depletion timescale
\ba
\left|\frac{d\log M_{\rm disk}}{dt}\right|^{-1}>~\tau_{\rm sec}^2\frac{{\phi}_{\rm out,disk}}{L\out}
\left[\frac{1}{I_{\rm out,0}}\right]^{4/3}\left[\frac{1}{1+\eta _\star}\right]^{2/3},
\ea
which it can be evaluated at the resonance encounter\footnote{It could also be evaluated at the time that the separatrix appears at $\Delta=1$, introducing a small correction.} $\Delta=0$ yields
\ba
\label{eq:disk_evolution_timescale}
\left|\frac{d\log M_{\rm disk}}{dt}\right|^{-1}>
\left[\frac{1}{I_{\rm out,0}}\right]^{4/3}\left(1+\eta _\star\right)^{1/3}
~\tau_{\rm sec}.
\ea
\end{enumerate}

Finally, we can compute the fixed points that describe the evolution of system. Using the canonical momentum-coordinate pair $(x,y)=\sqrt{2R}(\cos r, \sin r)$ we evaluate the fixed points of the Hamiltonian by setting $\partial\mathcal{K}/\partial x=0$, yielding:
\ba
x^3-3\Delta x-2=0.
\ea
For $\Delta<0$, when the disk dominates, there is only one branch with solution \citep{PMT2012}:
\ba
x^*(t)=\left(1+\sqrt{1-\Delta(t)^3}\right)^{1/3}+\Delta(t)\left(1+\sqrt{1-\Delta(t)^3}\right)^{-1/3}.
\ea
Thus, the (adiabatic) evolution of the system along the fixed point is simply given by $\sqrt{2R}=x^*(t)$ and $r=0$ ($\Omega\i-\Omega\out=\pi$, anti-aligned nodes).

\section{C. Unstable regions at high inclinations}\label{app:instability}

For simplicity  we assume an axisymmetric system with $\hat{\bf s}=\hat{\bf j}\out$ and ignore the disk that only allows to sweep over a range of inclinations $I\i$. In this limit, the Hamiltonian can be written in orbital elements as
\ba
\mathcal{H}=-\frac{\phi_{\rm in,out}}{2}\left(  -5e\i^2\sin I\i^2\sin^2 \omega\i+(1-e\i^2)\cos^2I\i +2e\i^2-\tfrac{1}{3}\right)-
\frac{\phi_{\rm in,\star}}{2(1-e\i^2)^{3/2}}\left(\cos^2I\i-\tfrac{1}{3}\right)
-\frac{\phi_{\rm in, GR}}{2(1-e\i^2)^{1/2}},
\ea
which we can write in terms of the Delaunay canonical variables as
\ba
\mathcal{H}=-\frac{\phi_{\rm in,out}}{2}\left[ \tfrac{5}{3}+
\frac{H\i^2}{L\i^2} -2\frac{G\i^2}{L\i^2}
-5\left(1-\frac{G\i^2}{L\i^2}-\frac{H\i^2}{G\i^2}+\frac{H\i^2}{L\i^2} \right)\sin^2\omega\i
\right]-
\frac{\phi_{\rm in,\star}}{2}\left(\frac{H\i^2L\i^3}{G\i^5}-\frac{L\i^3}{3G\i^3}\right)
-\frac{\phi_{\rm in,GR}L\i}{2G\i}.
\ea
From Hamilton's Equations $\dot{G}\i=-\partial \mathcal{H}/\partial \omega\i$ and $\dot{\omega}\i=\partial \mathcal{H}/\partial G\i$
\ba
\tau_{\rm sec}\dot{e}&=&5e\i(1-e\i^2)^{1/2}\sin^2I\i\sin \omega\i\cos\omega\i\\
\tau_{\rm sec}\dot{\omega}\i&=&
2(1-e^2)^{1/2}-5\left[(1-e\i^2)^{1/2}-\frac{\cos^2I\i}{(1-e\i^2)^{1/2}}\right]\sin^2\omega\i
+\frac{\eta _\star}{2(1-e\i^2)^2}\left(5\cos^2I\i-1\right)
+\frac{\eta_{\rm GR}}{2(1-e\i^2)},
\ea
with $\tau_{\rm sec}=L\i/\phi_{\rm in,out}$ and $H\i=L\i(1-e\i^2)^{1/2}\cos I\i$ a conserved quantity as $\mathcal{H}$ does not depend on $\Omega\i$. The linearized equations near the fixed point $e=0$ read
\begin{equation}
\frac{d}{dt}
\begin{pmatrix}
e\i\cos\omega\i\\ e\i\sin\omega\i
\end{pmatrix}
=\tau_{\rm sec}^{-1}
\begin{pmatrix}
0 & -A + B\\
A& 0
\end{pmatrix}
\begin{pmatrix}
e\i\cos\omega\i\\ e\i\sin\omega\i
\end{pmatrix}
\end{equation}
with $A=2+2\eta _\star+\eta_{\rm GR}/2-5/2\eta _\star\sin^2I\i $ and $B=5\sin^2 I\i$. We can then obtain the growth rates of the eccentricity vector by solving the eigenvalues of the square matrix as
\ba
\lambda&=&\pm\tau_{\rm sec}^{-1}\sqrt{A(B-A)}\nonumber\\
&=&\pm \tau_{\rm sec}^{-1} \left[
\left(2+2\eta _\star+\tfrac{1}{2}\eta_{\rm GR}-\tfrac{5}{2}\eta _\star\sin^2I\i\right) 
\times\left(5\sin^2I+\tfrac{5}{2}\eta _\star\sin^2I\i- 2-2\eta _\star-\tfrac{1}{2}\eta_{\rm GR}\right) \right]^{1/2}.
\ea
Thus, the fixed point $e\i=0$ is an unstable saddle point if the eigenvalues are real and different, requiring that $B>A>0$. Expressing this condition in terms of the inclinations, we get that the unstable range is given by
\ba
\left(\frac{4+4\eta _\star+\eta_{\rm GR}}{10+5\eta _\star}\right)<\sin^2I\i<\left(\frac{4+4\eta _\star+\eta_{\rm GR}}{5\eta _\star}\right).
\ea
We note that, for $\eta_{\rm GR}=0$, this expression is the same as the one found by \citet{katzdong2011} and \citet{TY2014} using the vectorial formalism without relativistic precession.

 

\end{document}